\documentclass[nohyper]{JHEP3}
\usepackage{amsmath,amssymb}
\usepackage{afterpage}
\usepackage{graphicx}
\usepackage{ifthen}
\usepackage{cite}
\usepackage{epstopdf}

\graphicspath{{Figs-EPS/}{Figs-PDF/}}

\makeatletter
\DeclareRobustCommand\recite[1]{\begingroup\@fileswfalse\cite{#1}\endgroup}
\makeatother

\newcommand{\Dmq}{\Delta m^2}

\def\lsim{\raise0.3ex\hbox{$\;<$\kern-0.75em\raise-1.1ex
\hbox{$\sim\;$}}}
\def\gsim{\raise0.3ex\hbox{$\;>$\kern-0.75em\raise-1.1ex
\hbox{$\sim\;$}}}
\DeclareMathAlphabet{\mathsc}{OT1}{cmr}{m}{sc}

\title{Robust Cosmological Bounds on Neutrinos and their Combination with
Oscillation Results}

\author{M.~C.~Gonzalez-Garcia\\
  C.N.~Yang Institute for Theoretical Physics\\
  State University of New York at Stony Brook\\
  Stony Brook, NY 11794-3840, USA,\\
  {\rm and:}
  Instituci\'o Catalana de Recerca i Estudis Avan\c{c}ats (ICREA),\\
  Departament d'Estructura i Constituents de la Mat\`eria and
  Institut de Ciencies del Cosmos,
  Universitat de Barcelona, Diagonal 647, E-08028 Barcelona, Spain\\
  E-mail:~\email{concha@insti.physics.sunysb.edu}}

\author{Michele Maltoni\\
  Instituto de F\'{\i}sica Te\'orica UAM/CSIC,
  Facultad de Ciencias, Universidad Aut\'onoma de Madrid,
  Cantoblanco, E-28049 Madrid, Spain\\
  E-mail:~\email{michele.maltoni@uam.es}}

\author{Jordi Salvado\\
  Departament d'Estructura i Constituents de la Mat\`eria and
  Institut de Ciencies del Cosmos,
  Universitat de Barcelona, 647 Diagonal, E-08028 Barcelona, Spain\\
  E-mail:~\email{jsalvado@ecm.ub.es}}

\abstract{
We perform a global analysis of cosmological observables in
generalized cosmologies which depart from 
$\Lambda$CDM models by allowing non-vanishing curvature  
$\Omega_k\neq 0$,  dark energy with equation of state with 
$\omega\neq -1$, the presence of additional relativistic degrees of freedom
$\Delta N_{\rm rel}$, and neutrino
masses $\Omega_\nu\neq 0$.  By combining the data from
cosmic microwave background (CMB) experiments (in particular the
latest results from WMAP-7), the present day Hubble constant (H0) 
measurement, the high-redshift Type-I supernovae (SN) results and the 
information
from large scale structure (LSS) surveys, 
we determine the parameters
in the 10-dimensional parameter space for such models. We present the
results from the analysis when the full shape information from the LSS
matter power spectrum (LSSPS) is included versus when only the corresponding
distance measurement from the baryon acoustic oscillations (BAO) is
accounted for. We compare the bounds on the neutrino mass scale in
these generalized scenarios with those obtained for the 6+1 parameter
analysis in $\Lambda{\rm CDM}+m_\nu$ models and we also study the
dependence of those on the set of observables included in the
analysis.  Finally we combine these results with the information on
neutrino mass differences and mixing from the global analysis of
neutrino oscillation experiments and derive the presently allowed
ranges for the two laboratory probes of the absolute scale of 
neutrino mass: the effective electron neutrino mass in single beta decay and 
the effective Majorana neutrino mass in neutrinoless $\beta\beta$ decay.}

\preprint{YITP-SB-10-22\\IFT-UAM/CSIC-10-42}

\begin{document}

\section{Introduction}
It is now an established fact that neutrinos are massive and leptonic
flavors are not symmetries of Nature~\cite{Pontecorvo:1967fh,
Gribov:1968kq}.  In the last decade this picture has become fully
proved thanks to the upcoming of a set of precise experiments. In
particular, the results obtained with 
solar~\cite{Cleveland:1998nv,Kaether:2010ag,Abdurashitov:2009tn,
Hosaka:2005um,Aharmim:2007nv,Aharmim:2005gt,Aharmim:2008kc,Collaboration:2009gd,Arpesella:2008mt,Collaboration:2008mr}
and atmospheric neutrinos~\cite{Ashie:2005ik,Wendell:2010md}
have been confirmed in experiments using terrestrial beams: neutrinos
produced in nuclear reactors~\cite{Shimizu:2008zz,CHOOZ}
and accelerators~\cite{Ahn:2006zza,Adamson:2008zt,Collaboration:2009yc,minapp70} facilities  have been detected at distances of the order of hundreds of
kilometers~\cite{ourrep}.

The minimum joint description of all the neutrino data requires mixing
among all the three known neutrinos ($\nu_e$, $\nu_\mu$, $\nu_\tau$),
which can be expressed as quantum superpositions of three massive
states $\nu_i$ ($i=1,2,3$) with masses $m_i$.  This implies the
presence of a leptonic mixing matrix in the weak charged current
interactions~\cite{Maki:1962mu, Kobayashi:1973fv} which can be
parametrized as:
\begin{equation}
    U =
    \begin{pmatrix}
	1 & 0 & 0 \\
	0 & c_{23}  & {s_{23}} \\
	0 & -s_{23} & {c_{23}}
    \end{pmatrix}
    \cdot
    \begin{pmatrix}
	c_{13} & 0 & s_{13} e^{-i\delta_\text{CP}} \\
	0 & 1 & 0 \\
	-s_{13} e^{i\delta_\text{CP}} & 0 & c_{13}
    \end{pmatrix}
    \cdot
    \begin{pmatrix}
	c_{21} & s_{12} & 0 \\
	-s_{12} & c_{12} & 0 \\
	0 & 0 & 1
    \end{pmatrix}
    \cdot
    \begin{pmatrix}
	e^{i \eta_1} & 0 & 0 \\
	0 & e^{i \eta_2} & 0 \\
	0 & 0 & 1
    \end{pmatrix},
    \label{eq:U3m}
\end{equation}
where $c_{ij} \equiv \cos\theta_{ij}$ and $s_{ij} \equiv
\sin\theta_{ij}$.  In addition to the Dirac-type phase
$\delta_\text{CP}$, analogous to that of the quark sector, there are
two physical phases $\eta_i$ associated to the Majorana character of neutrinos
and which are not relevant for neutrino
oscillations~\cite{Bilenky:1980cx, Langacker:1986jv}.  

Given the observed hierarchy between the
solar and atmospheric mass-squared splittings there are two possible
non-equivalent orderings for the mass eigenvalues, which are
conventionally chosen as
\begin{align}
\label{eq:normal}
m_1< m_2< m_3 \;\;\;  {\rm with} \;\;\; 
  \Dmq_{21} &\ll (\Dmq_{32} \simeq \Dmq_{31})
  \text{ with } (\Dmq_{31} > 0) \,;
  \\
  \label{eq:inverted}
m_3< m_1< m_2 \;\;\;  {\rm with} \;\;\; 
  \Dmq_{21} &\ll |\Dmq_{31} \simeq \Dmq_{32}|
  \text{ with } (\Dmq_{31} < 0) \,.
\end{align}
As it is customary we refer to the first option,
Eq.~\eqref{eq:normal}, as the \emph{normal}  (N) scheme, and to the second
one, Eq.~\eqref{eq:inverted} , as the \emph{inverted} (I) scheme; in
this form they correspond to the two possible choices of the sign of
$\Dmq_{31}$.  In this convention the angles $\theta_{ij}$ can be taken
without loss of generality to lie in the first quadrant, $\theta_{ij}
\in [0, \pi/2]$, and the phases $\delta_\text{CP},\; \eta_i\in [0,2\pi]$.

Within this context, $\Dmq_{21}$, $|\Dmq_{31}|$, $\theta_{12}$, and
$\theta_{23}$ are relatively well
determined from oscillation experiments
~\cite{ourfit,Fogli:2009zza,Schwetz:2008er,Maltoni:2008ka}, 
while only an upper bound is derived for the mixing
angle $\theta_{13}$  and barely nothing is known on the  phases
and on the sign of $\Dmq_{31}$. Furthermore neutrino oscillation data   
provides as unique information on the absolute neutrino mass scale a lower 
bound 
\begin{eqnarray}
&&\sum m_i \gtrsim \sqrt{|\Delta m^2_{31}|}\;\; 
{\rm for}\; {\rm N}\\
&&\sum m_i \gtrsim 2 \sqrt{|\Delta m^2_{31}|} \;\; 
{\rm for}\; {\rm I}\\
\end{eqnarray}

Conversely the neutrino mass scale is constrained in laboratory experiments 
searching for its kinematic effects in  Tritium $\beta$ decay which 
are sensitive to the so-called effective electron neutrino mass~\cite{betamix,vissanitrit,smirtrit}
\begin{equation}
    m^2_{\nu_e} \equiv 
    \sum_i m^2_i |U_{ei}|^2=c_{13}^2 c_{12}^2 m_1^2
+c_{13}^2 s_{12}^2 m_2^2+s_{13}^2 m_3^2 \,,
    \label{eq:mb}
\end{equation}
At present the most precise determination from the
Mainz~\cite{tritmainz} and Troitsk~\cite{trittroitsk} experiments 
give no indication in favor of $m_{\nu_e}\neq 0$ and one sets an
upper limit 
\begin{equation}
    \label{eq:nuelim}
    m_{\nu_e}<2.2~\text{eV} \; , 
\end{equation}
at 95\% confidence level (CL).  
A new experimental project, KATRIN~\cite{katrin}, is under
construction with an estimated sensitivity limit: $m_{\nu_e} \sim 0.2$
eV.

Direct information on neutrino masses can also be obtained from 
neutrinoless double beta decay ($0\nu\beta\beta$) searches provided
they are Majorana particles. In the absence of other sources of 
lepton number violation in the low energy lagrangian, 
the $0\nu\beta\beta$ decay amplitude is proportional to the
effective Majorana mass of $\nu_e$, $m_{ee}$,
\begin{equation}
    m_{ee} = \left| \sum_i m_i U_{ei}^2 \right|=
\left |
c_{13}^2 c_{12}^2 m_1\, {e}^{i\eta_1} +
c_{13}^2 s_{12}^2 m_1\, {e}^{i\eta_2} +
s_{13}^2\, {e}^{-i\delta_{CP}}  \;, 
\right|
    \label{eq:mbb}
\end{equation}
which, in addition to the masses and mixing parameters that affect the
tritium beta decay spectrum, depends also on the phases in the leptonic
mixing matrix. The strongest bound from $0\nu\beta\beta$ decay was 
imposed by the  Heidelberg-Moscow group~\cite{hmlimit} 
\begin{equation}
    m_{ee} < 0.26~(0.34)~\text{eV}
    \quad \text{at 68\% (90\%) CL,}
\end{equation}
which holds for a given 
prediction of the nuclear matrix element. However, there are large 
uncertainties in those predictions which may 
considerably weaken the bound~\cite{bbteoreview}.
A series of new experiments is planned with sensitivity of up to
$m_{ee} \sim 0.01$ eV~\cite{bbreview}.  

Neutrinos, like any other particles, contribute to the total energy
density of the Universe.  Furthermore within what we presently know 
of their masses,  the three Standard Model (SM)  neutrinos are relativistic 
through most  of the evolution of the Universe and they are very weakly 
interacting which means that they decoupled early in cosmic history. 
Depending on their exact masses they can impact the CMB spectra, 
in particular  by  altering the value of the redshift for matter-radiation 
equality. 
More importantly, their free  streaming suppresses  the growth of structures 
on scales smaller than the horizon at the time when 
they become non-relativistic  and therefore affects the matter
power spectrum which is probed from surveys of the LSS distribution 
(see \cite{pastor} for a detailed review of cosmological effects of neutrino 
mass).
 
Within their present precision, cosmological observations are sensitive 
to neutrinos only via their contribution to the energy 
density in our Universe, $\Omega_\nu h^2$ (where $h$ is the Hubble 
constant normalized to $H_0 = 100 ~\text{km} ~\text{s}^{-1}
~\text{Mpc}^{-1}$). $\Omega_\nu h^2$ is related to the
total mass in the form of neutrinos 
\begin{equation} 
    \Omega_{\nu}h^2 = \sum_i m_i / (94 \text{eV}) \,.
\end{equation}
Therefore cosmological data mostly gives information on the sum of the
neutrino masses and has very little to say on their mixing structure 
and on the ordering of the mass states
(see Ref. \cite{jkpv} for a recent update on the sensitivity of 
future cosmological observations to the mass ordering.)

There is a growing literature on the information extracted from cosmological
observations on the neutrino mass scale, starting with the analysis
performed by the different experimental 
collaborations~\cite{WMAP7,WMAP5,WMAP3,BAO,SDSS}. 
The basic observation is that, besides variations due to the observables
considered,  the bounds on the neutrino mass  obtained depend on the 
assumptions made on the history of the cosmic expansion, or in other words, 
on how many parameters besides the $\Lambda$CDM model are  allowed to vary 
when analyzing the cosmological data.  Additionally depending on those 
assumptions some observables need or not to be considered in order
to account for degeneracies among the parameters 
(for some recent analysis  see ~\cite{hannestad,hamann,reid}).  

In this article we present the results of a global analysis of cosmological 
observables in $o\omega{\rm CDM}+\Delta N_{\rm rel}+m_\nu$
cosmologies which depart from  $\Lambda$CDM models by allowing, 
besides neutrino masses $\Omega_\nu\neq 0$,   
non-vanishing curvature   $\Omega_k\neq 0$, dark energy 
with equation of state with $\omega\neq -1$ together with 
the presence of new particle physics whose effect on the present 
cosmological observations can be  parametrized in terms of additional 
relativistic degrees of freedom $\Delta N_{\rm rel}$. In particular 
this extends the most general analysis of Ref.~\cite{hamann} 
by accounting  also for non-flatness effects.  We adopt a purely 
phenomenological approach in analyzing  the effect of a non-vanishing
spatial curvature without addressing its origin. However
it is worth mentioning that, within inflationary models which produce
the simple initial conditions here considered it is difficult to end up
with  a significant $\Omega_k$ \cite{infla}.
We describe in Sec.~\ref{sec:inputs} the different cosmological
observables included in these 10-parameter analysis as well as 
our statistical treatment of those. The results of the analysis
are presented in Sec.~\ref{sec:cosmoana} where we discuss 
the differences obtained when  the full shape information from the LSS
matter power spectrum is included versus when only the corresponding
distance measurement from the baryon acoustic oscillations  is
accounted for. We also  compare the bounds on the neutrino mass scale in
these $o\omega{\rm CDM}+\Delta N_{\rm rel}+m_\nu$
scenarios with those obtained for the 6+1 parameter
analysis in $\Lambda{\rm CDM}+m_\nu$ models and we also study the
dependence of those on the set of observables included in the
analysis.  These results are combined with  the information on
neutrino mass differences and mixing from the global analysis of
neutrino oscillation experiments in Sec.\ref{sec:mbb} to 
derive the presently allowed ranges 
for the two laboratory probes of the 
absolute scale of neutrino mass: the effective neutrino mass in single 
beta decay $m_{\nu_e}$ and  the effective Majorana neutrino mass 
in neutrinoless  $\beta\beta$ decay $m_{ee}$. 
We summarize our conclusions in Sec.\ref{sec:conclu}.

\section{Cosmological Inputs and Data Analysis}
\label{sec:inputs}
\TABLE{
\begin{tabular}{|l|c|}
\hline
Parameter & symbol\\
\hline
Hubble Constant & $H_0$ \\
Baryon density & $\Omega_b h^2$ \\
Dark matter density & $\Omega_c h^2$\\
Scalar spectral index & $n_s$ \\
Optical Depth at Reonization  & $\tau$\\
Amplitude of scalar power spectrum at $k=0.05$ Mpc$^{-1}$& $A_S$ \\
\hline
Total neutrino mass  & ${\displaystyle \sum_{i=1,3} m_{\nu,i}} $ \\
Dark energy equation of state parameter & $\omega$ \\
Effective  number of 
extra relativistic degrees of freedom & 
$\Delta N_{\rm rel}$\\
Spatial curvature density & $\Omega_k$ \\
\hline
\end{tabular}
\caption{Cosmological parameters used in our most general analysis.  
$h=H_0/100$. We denote the cosmology characterized by these parameters 
as $o\omega{\rm CDM}+\Delta N_{\rm rel}+m_\nu$.}
\label{tab:param}
}
We consider cosmologies  $o\omega{\rm CDM}+\Delta N_{\rm rel}+m_\nu$
characterized by the free parameters listed in 
Table ~\ref{tab:param}. All parameters are as usually defined in the
literature with the exception of $\Delta N_{\rm rel}$. 
Our definition of extra relativistic degrees of freedom
accounts for the fact that we have evidence of the existence of three
and only three standard neutrino species which mix due to mass
oscillations \cite{ourrep}. 
Their contribution to the energy budget of the universe
is included in $\Omega_\nu h^2$.  $\Delta
N_{\rm rel}$ parametrizes the contribution 
of additional relativistic massless states of any spin 
to the radiation energy density. 
For convenience that contribution is normalized to the one from a spin
$1/2$ weakly interacting massless state. That normalization is $\Delta
N_{\rm rel}$. 

In these cosmologies several parameter degeneracies appear in any of
the cosmological observables. First, any experiment that measures the
angular diameter or luminosity distance to a single redshift is not
able to constrain $\Omega_k$  because the distance depends not
only on $\Omega_k$, but also on the expansion history of the
universe. Thus for a universe containing matter and vacuum energy, one
needs to combine at least two absolute distance indicators, or the
expansion rates out to different redshifts to break this degeneracy.
Furthermore when dark energy is dynamical, $\omega\neq-1$ , a
third distance indicator is required. Finally the presence of extra 
relativistic degrees of freedom $\Delta N_{\rm rel}$ changes the 
matter-radiation equality epoch, a change that can be compensated by the 
corresponding
modification of the matter density $\Omega_m h^2$.  As a result,
$\Delta N_{\rm rel}$ and $\Omega_mh^2$ are strongly degenerate unless
a fourth distance indicator provides us with an independent constraint
on $\Omega_mh^2$.  

In our analysis we include the results from the 7-year data of WMAP
\cite{WMAP7} on the temperature and polarization anisotropies in the
form of the temperature (TT), E-mode polarization (EE), B-mode
polarization (BB), and temperature-polarization cross-correlation (TE)
power spectra for which we use the likelihood 
function as provided by the collaboration \footnote{We notice that, 
although the  models considered do not generate any B-mode polarization,  
in order to account for the information from EE and low-l TE data, 
the BB power spectrum must also  be included in the analysis because WMAP  
provides the combined  likelihood for the low-l TE, EE and BB 
spectra~\cite{larson}}.  
A number of CMB experiments have probed smaller
angular scales than WMAP. In particular we consider the results from
the temperature power spectra from the Cosmic Background Imager (CBI)
\cite{CBI}, the Very Small Array (VSA)\cite{VSA}, BOOMERANG
\cite{BOOMERANG} and the Arcminute Cosmology Bolometer Array Receiver
(ACBAR), \cite{ACBAR}.  In order to avoid redundancies among the CMB
data sets, we follow the procedure in Refs.\cite{WMAP5,finelli}.  We
use seven band powers for CBI (in the range $948<\ell<1739$), five for
VSA ($894<\ell<1407$), seven for BOOMERANG ($924<\ell<1370$), and sixteen
band powers of ACBAR in the range $900<\ell<2000$. We do not include
the results of the Background Imaging of Cosmic Extragalactic
Polarization (BICEP) \cite{BICEP} experiment whose bands overlap
excessively with WMAP and from QUaD \cite{QUAD} which observes the same
region of sky as ACBAR and it is less precise \cite{finelli}.
Furthermore we do not include in the analysis the polarization results
of these experiments. As mentioned above, in the analysis of WMAP we use the 
likelihood 
function as provided by the collaboration. For the other CMB experiments 
we build the corresponding likelihood functions from the data, covariance 
matrix and window functions given by each experiment. 
We compute theoretical CMB predictions 
using the fast Boltzmann code CAMB~\cite{CAMB,CMBFAST}. 
Following the procedure outlined in Ref.\cite{WMAP3}
whenever it is required we account for the Sunyaev-Zel'dovich (SZ)
effect by marginalizing over the amplitude of the SZ contribution
parametrized by the model of Ref.\cite{KS}. We assume a uniform prior
on the amplitude as $0< A_{\rm SZ}< 2$.

We also include the results from Ref.\cite{H0} on the present-day
Hubble constant, $H_0=74.2\pm 3.6~{\rm km~s^{-1}~Mpc^{-1}}$ where the
quoted error includes both statistical and systematic errors.  This
measurement of $H_0$ is obtained from the magnitude-redshift relation
of 240 low-$z$ Type Ia supernovae at $z<0.1$. We include this result
as a Gaussian prior and neglect the slight cosmology dependence \cite{reid}
of this constraint. 

The results from luminosity measurements of  high-$z$ Type Ia
supernovae are included as presented in the compilation of the
supernova data called the ``Constitution'' sample in Ref.\cite{SN09}
which consists of 397 supernovae and it is an extension of the
``Union'' sample \cite{SN08}.  With these data we build the
corresponding likelihood function without including systematic errors
whose precise values are still under debate \cite{SN09}. In our
analysis we marginalize over the absolute magnitude of the supernovae
with a uniform prior.

Finally we also include the results from the matter power spectrum 
as derived from large scale structure surveys in two different 
forms. In one  case we use the measurement of BAO scale obtained 
from the Two-Degree Field Galaxy Redshift Survey (2dFGRS) and 
the Sloan Digital Sky Survey Data Release 7 (SDSS DR7) 
\cite{BAO}. In the other we include the full 
power spectrum  of the SDSS DR7 survey \cite{SDSS} (which 
we label LSSPS). 

For the analysis including the BAO scale, we use as input data 
the two distance ratios at $z=0.2$ and $z=0.35$ presented in Ref.\cite{BAO} 
and build the corresponding
likelihood function using the covariance matrix as given in that reference.
As discussed in Ref.\cite{BAO} the distance ratios can be considered
as measurements of $d_z \equiv rs(z_d)/D_V(z)$ and apply to any
of the considered models. $r_s(z_d)$ is
the comoving sound horizon at the baryon drag epoch and $D_V(z) =
[(1+z)^2 D_A^2 c z/H(z)]^{1/3}$ with $D_A$ is the angular diameter
distance and $H(z)$ is the Hubble parameter.
However in their fitting procedure the value of $d_z$ is
obtained by first assuming some fiducial cosmology, extracting the
value of $D_V(z)$ and then computing $rs(z_d)
/ D_V(z)$ with $r_s(z_d)$ evaluated by that fiducial cosmology
using the approximated formula of  Eisenstein \& Hu \cite{z_d} for $z_d$.
As discussed in Ref.\cite{hamann} this approximated formula 
is not strictly valid for the  extended cosmologies which we are considering. 
We correct for this effect by a) exactly evaluating the redshift at 
baryon drag epoch by using Eq.(B.5) in Ref.\cite{hamann} 
in the extended cosmologies, and b) correcting for the use of the 
approximate formula in the presentation of the data by rescaling
the predictions by a factor  
$r^{\rm fid}_s(z_{d\, \rm approx})/r^{\rm fid}_s(z_{d\, \rm exact})$
(we prefer to rescale the predictions since the
covariance matrix is given for the data as presented). 

In our second analysis we include the full SDSS DR7 data which
consists of 45 bins, covering wavenumbers from $k_{\rm min} =
0.02\ h {\rm Mpc}^{-1}$ to $k_{\rm max} = 0.2\ h {\rm Mpc}^{-1}$
(where $k_{\rm min}$ and $k_{\rm max}$ denote the wavenumber at which
the  window functions of the first and last data point have their
maximum). In this analysis we use the likelihood function as provided
by the experiment. Together with the  linear matter power spectrum
it requires a smooth version of it with the baryon oscillations
removed.  We construct such no-wiggle spectrum for the extended cosmologies 
here considered from the linear matter power spectrum computed by CAMB
using the method based on the discrete spectral analysis of the power
spectrum described in Appendix A.1 of Ref.\cite{hamann}.

We also perform comparative analysis including only the $\Lambda$CDM
parameters plus massive neutrinos (first seven in Table~\ref{tab:param})
fixing $\omega=-1$ $\Delta N_{\rm rel}=\Omega_k=0$ for different combinations
of the above observables. 

Additional constraints on the cosmological parameters can be
obtained if one includes in the analysis information on
the growth of structure from other low redshift data. Among others the small scale
primordial spectrum determined from Lyman-alpha forest clouds or the
priors on the amplitude of mass fluctuations derived from different 
galaxy cluster samples.
We have conservatively chosen not to include those in our analysis because 
generically these results are subject to model dependence assumptions 
which render them not directly applicable for the most general cosmologies
here consider.

With the data from the different samples included  in  a given analysis 
and the theoretical predictions for them in terms of the relevant 
parameters $\vec x$, 
we construct the corresponding combined likelihood function.
In Bayesian statistics our knowledge of $\vec x$ is summarized by
the posterior probability distribution function (p.d.f.)
\begin{equation}
    p(\vec x|\mathrm{D},\mathcal{P}) =
    \dfrac{\mathcal{L}(\mathrm{D} | \vec x)\, \pi(\vec x | \mathcal{P})}
    {\int \mathcal{L}(\mathrm{D} | \vec x')\, \pi(\vec x' | \mathcal{P})\, d\vec x'} \,.
    \label{eq:ppdf}
\end{equation}
$\pi(\vec x | \mathcal{P})$ is the prior probability density for 
the parameters. In our analysis we assume a uniform
prior probability for the $\vec x$  parameters in Table \ref{tab:param}.
For $\sum m_{\nu}$ and $\Delta N_{\rm rel}$ we imposed that they 
should be both $\geq 0$. 
Following standard techniques in order to 
reconstruct the posterior p.d.f.\
Eq.~\eqref{eq:ppdf} we have developed a  Markov Chain Monte Carlo  
(MCMC) generator 
which employs the Metropolis-Hasting algorithm including
the adapting for the kernel function to increase the efficiency.  
Full details are given in Appendix B of Ref.\cite{oursolar}.
For each combination of data we generate ${\cal O} (50)$ chains 
in parallel and verify its  convergence by studying the variation of the 
Gelman-Rubin $R$-parameter \cite{rubin}
imposing as convergence criteria $R-1\lesssim 5\times 10^{-2}$. 

\FIGURE[!h]{
\includegraphics[width=0.8\textwidth]{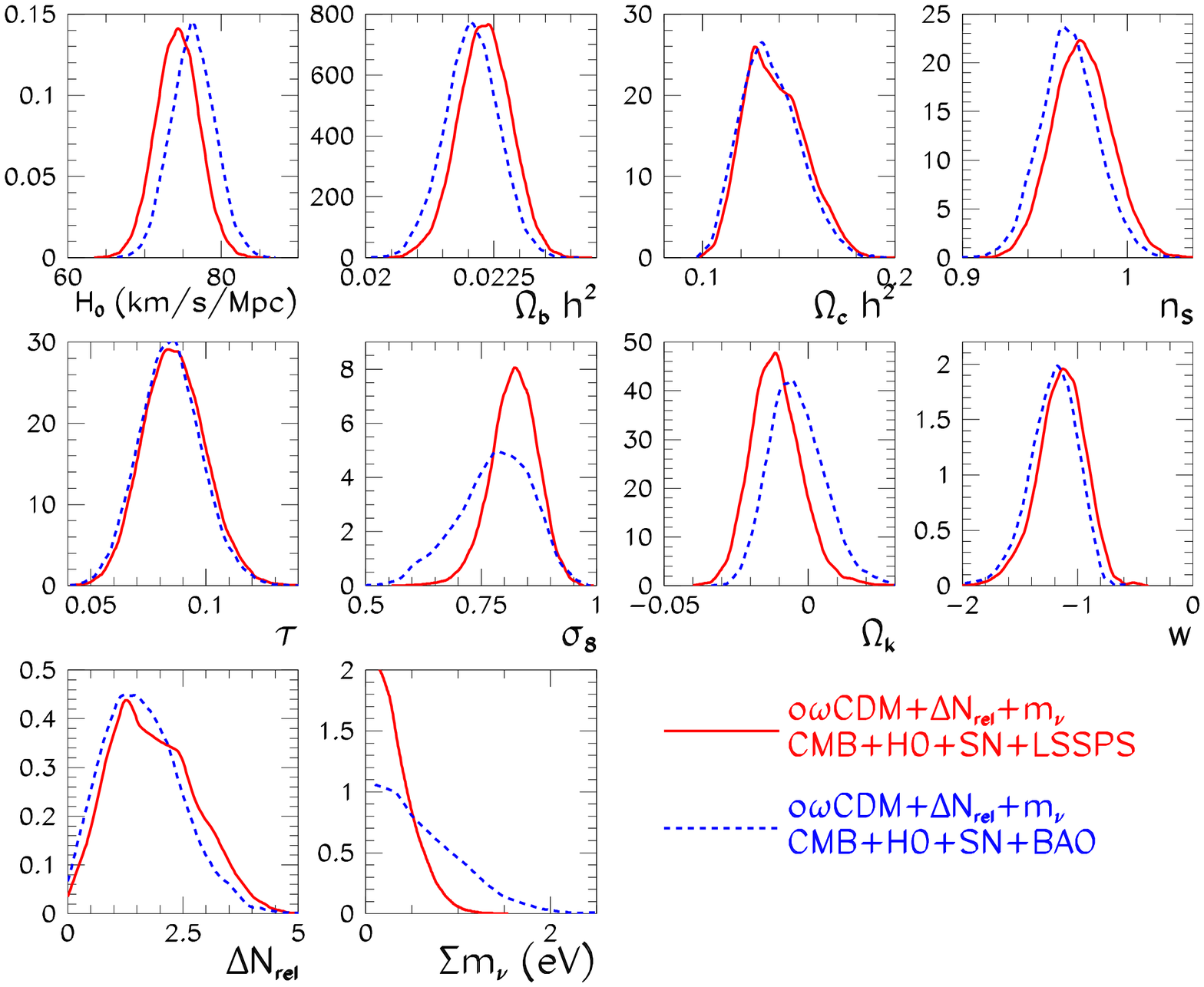}
 \caption{
    Constraints from our global analysis on the cosmological parameters
of $o\omega{\rm CDM}+\Delta N_{\rm rel}+m_\nu$
    for the analysis including CMB+H0+SN+LSSPS (solid red)  and 
   CMB+H0+SN+BAO (dotted blue). The different panels
    show the marginalized one-dimensional probability distributions 
    for all parameters. For  the neutrino mass see also Fig.\ref{fig:mnu}.}
\label{fig:1dim}
}

\TABLE{
\begin{tabular}{|c||c|c|c||c|c|c|} 
\hline
 &\multicolumn{3}{c|} {CMB+HO+SN+BAO} &
 \multicolumn{3}{c|} {CMB+HO+SN+LSS-PS} \\ 
\hline
& best & 1$\sigma$ & 95\% CL 
& best & 1$\sigma$ & 95\% CL  \\[+0.2cm]
\hline 
&&&&&&
\\[-0.3cm]
$H_0$ km/s/Mpc 
& 76.2 & $^{ + 3.0} _ {- 2.8}$ & $^{ + 5.7} _  {- 5.6} $  
& 74.4 & $^{ + 2.8} _ {- 2.9}$ & $^{ + 5.6} _  {- 5.6} $  
\\[+0.2cm]
$\Omega_b h^2\times 100$
& 2.205 & $^{ + 0.057} _ {- 0.050} $ & $^{ + 0.103} _  {- 0.105} $ 
& 2.239 & $^{ + 0.059} _ {- 0.046} $  &$^{ + 0.095} _  {- 0.108} $  
\\ [+0.2cm]
$\Omega_c h^2$ 
& 0.131 & $^{ + 0.018} _ {- 0.013} $ & $^{ + 0.036} _  {- 0.023} $
&0.128 & $^{ + 0.024} _ {- 0.009} $ &$^{ + 0.042} _  {- 0.018} $ 
\\[+0.2cm]
$n_S$ 
&  0.961&$^{ + 0.021} _ {- 0.015} $&$^{ + 0.040} _  {- 0.030} $ 
&  0.971&$^{ + 0.019} _ {- 0.017} $&$^{ + 0.037} _  {- 0.033} $ 
\\ [+0.2cm]
$\tau$ 
&   0.086& $^{ + 0.011} _ {- 0.015}$ & $^{ + 0.026} _  {- 0.028} $ 
&  0.083& $^{ + 0.016} _ {- 0.011}$ &  $^{ + 0.030} _  {- 0.023} $ 
\\[+0.2cm]
$\sigma_8$ 
& 0.787& $^{ + 0.091} _ {- 0.073} $& $^{ + 0.135} _  {- 0.179} $ 
&0.824& $^{ + 0.051} _ {- 0.048} $ &$^{ + 0.097} _  {- 0.105} $ 
\\ [+0.2cm]
$\Omega_k$ 
&  -0.006& $^{ + 0.010} _ {- 0.009} $ & $  -0.022\leq \Omega_k\leq 0.016   $ 
& -0.011&  $^{ + 0.008} _ {- 0.009} $ & $  -0.028\leq \Omega_k\leq 0.007   $ 
\\[+0.2cm]
$\omega$ 
&   -1.17&  $^{ + 0.19} _ {- 0.21} $ & $  -0.62\leq\omega+1\leq 0.18    $ 
&  -1.12 &  $^{ + 0.21} _ {- 0.20} $ & $  -0.57\leq\omega+1\leq 0.26    $ 
\\[+0.2cm]
$\Delta N_{\rm rel}$
&1.2& $^{ + 1.1} _ {- 0.61} $ & $0.08\leq \Delta N_{\rm rel}\leq 3.2$ 
&1.3& $^{ + 1.4} _ {- 0.54} $ & $0.21\leq \Delta N_{\rm rel}\leq 3.6$ 
\\[+0.2cm]
$\sum m_\nu$ (eV) 
& &$\leq 0.77$ & $\leq 1.5$  
& & $\leq 0.37$ &  $\leq 0.76$
\\
\hline
\end{tabular}
\caption{Constraints from our global analysis for 
$o\omega{\rm CDM}+\Delta N_{\rm rel}+m_\nu$ cosmologies.  
We show the values for the best fit parameters and 
the corresponding 1$\sigma$ (68\%) and 2$\sigma$ (95\%) allowed intervals.}
\label{tab:1dim}
}

\FIGURE[!h]{
\includegraphics[height=0.7\textheight]{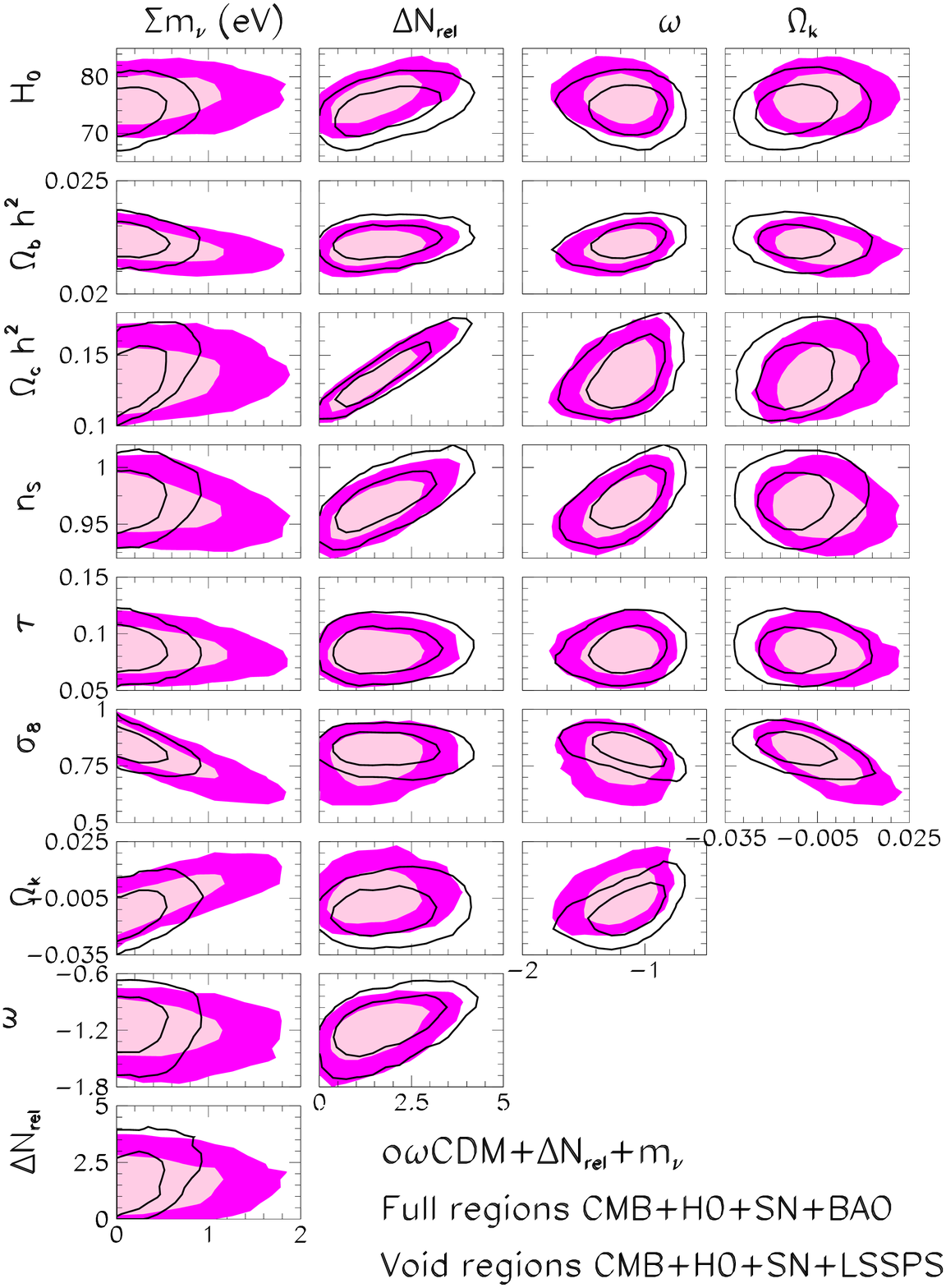}
  \caption{Constraints from our global analysis for 
$o\omega{\rm CDM}+\Delta N_{\rm rel}+m_\nu$ cosmologies  
    for the analysis including CMB+H0+SN+BAO (full regions) and 
for for the analysis including CMB+H0+SN+LSSPS (void regions).
    We show the 68\% and 95\% CL two-dimensional credibility regions 
    for the last four parameters in Table \ref{tab:param}}
\label{fig:fit10}
}

\section{Results of the Cosmological Fits}
\label{sec:cosmoana}
Our results for the two analysis in 
$o\omega{\rm CDM}+\Delta N_{\rm rel}+m_\nu$ cosmologies are presented 
in  Figs.~\ref{fig:1dim}--~\ref{fig:fit10} and in Table \ref{tab:1dim}.
In Fig.~\ref{fig:1dim} we show the marginalized one-dimensional probability 
distributions for the ten independent parameters obtained from 
Eq.\eqref{eq:ppdf} as 
\footnote{Technically this is obtained from the MCMC chain by discretizing the 
parameter space and counting the fraction of points in each cell.}
\begin{equation}
p_{\rm 1-dim}(x_i) =\int dx_{k\neq i} \, p(\vec x|\mathrm{D},\mathcal{P}) \; . 
\end{equation}
For convenience we show the information on the normalization of the
scalar power spectrum in terms of the derived $\sigma_8$ parameter  which 
parametrizes the expected root mean square amplitude of the matter fluctuations
in spheres of radius $R=8 h^{-1}$ Mpc. 

The best fit values given in the second and fourth columns in 
Table \ref{tab:1dim} are those for which $p_{\rm 1-dim}(x^{\rm best}_i)$ 
is maximum. The allowed ranges at a given CL,  
$x^{\rm CL}_{i,\rm min}\leq x_i\leq x^{\rm CL}_{i,\rm max}$,  
are obtained from the condition
\begin{eqnarray}
&& {\rm CL} 
\left[x^{\rm CL}_{i,\rm min}\leq x_i\leq x^{\rm CL}_{i,\rm max}\right]
=\int_{x^{\rm CL}_{i,\rm min}}^{x^{\rm CL}_{i,\rm max}}
p_{\rm 1-dim}(x_i)  \label{eq:1dimrange} \\
&& {\rm with}  \;\;\;
p_{\rm 1-dim}(x^{\rm CL}_{i,\rm  min})=p_{\rm 1-dim}(x^{\rm CL}_{i,\rm max}) 
\label{eq:2side}\\
&&{\rm or} \;\;\; x^{\rm CL}_{i, \rm min}=0 \;\;\;{\rm for}\;\; 
x_i=\Delta N_{rel}, \sum  m_\nu \label{eq:1side}
\end{eqnarray}
where \eqref{eq:1side} is used when there is no solution for 
condition \eqref{eq:2side}. 

Equivalently we define the marginalized two-dimensional  probability 
distribution functions 
\begin{equation}
p_{\rm 2-dim}(x_i,x_j) 
=\int dx_{k\neq i,j} \, p(\vec x|\mathrm{D},\mathcal{P}) \; ,
\end{equation}
and from these, we obtain the two-dimensional credibility regions with a 
given CL as the region with smallest area and with CL integral 
posterior probability. In practice they are obtained as the regions
surrounded by a two-dimensional isoprobability contour which contains 
the point of highest posterior probability and within which the
integral posterior probability is CL. We plot in Fig.\ref{fig:fit10} 
the 68\% and 95\% CL two-dimensional credibility regions 
for the last four parameters in Table \ref{tab:param}.  
For the analysis 
including CMB+H0+SN+BAO (full regions) 
and including CMB+H0+SN+LSSPS (void regions). 

Because of the degeneracies present in these cosmologies one finds, as
expected, a degradation in the constraints of the {\sl standard}
parameters (ie those of the $\Lambda$CDM model) when compared with the
analysis performed within  the $\Lambda$CDM priors for the same set of
observables (see for example table 1 in Ref.~\cite{WMAP7}). As seen in
the figure this particularly affects the determination of $\Omega_c$
(or equivalently $\Omega_m$) as a consequence of the well--known
degeneracy in the predictions of the CMB spectra between $\Delta
N_{\rm rel}$ and $\Omega_m$. This is so because a simultaneous change of
both can leave untouched the redshift for matter-radiation equality, 
which is well constrained by the ratio of height of the third and first 
peaks in the
CMB spectra.  This degeneracy is broken by the addition of the H0
prior as well as the independent determination of $\Omega_m$ from the
distance information from LSS, either using only BAO or the full power
spectrum. It is interesting to notice that we find that 
data is better described by allowing a non-zero amount of extra radiation
even though it is only at most a 2$\sigma$ effect. This implies, for example,
that models with extra light sterile neutrinos are favoured by 
the data (as discussed in Ref.\cite{steriles} in the context of
flat cosmologies with a cosmological constant) even for these 
$o\omega$CDM models.   Most conservatively we can
read the results as a 2$\sigma$ upper bound on 
$\Delta N_{\rm rel}\leq 3.2$ (3.6) for the analysis 
including CMB+H0+SN+BAO (CMB+H0+SN+LSSPS). 

We also find a widening in the allowed range of $n_S$ as a consequence
of its degeneracy with $\Delta N_{\rm rel}$ and with the dark energy
equation of state $\omega$. Both change the ratio of generated
power at low versus large angular scales in the CMB spectrum, an
effect that can be offset by a change in the spectral index. 
Conversely we find that in these cosmologies 
$\omega$ is considerably less constrained than  in $\omega$CDM scenarios 
for which a 95\% range $-0.089\leq\omega+1\leq 0.12$ is obtained from the 
analysis of  CMB+BAO+SN ~\cite{WMAP7}.

The normalization of the power spectrum as parametrized by $\sigma_8$
is mostly affected by the presence of neutrino masses. Their main
effect is to reduce the amplitude on the power spectrum on
free-streaming scales therefore decreasing $\sigma_8$. There is also a
residual degeneracy between $\sigma_8$ and $\Omega_k$ mostly
associated with the fact that allowing for non-flat cosmologies
permits to increase the amount of dark matter in the form of
neutrinos without affecting $\Omega_c$ and therefore minimizing their
indirect impact on the CMB spectra.  As a consequence we find that
there is a correlation between the allowed range of neutrino mass and
$\Omega_k$ as it is seen in the corresponding panel of
Fig.\ref{fig:fit10}. This leads to a somewhat wider allowed range 
of $\Omega_k$ when compared to the results obtained  from the 
analysis of  CMB+BAO+SN in o$\omega$CDM scenarios 
$-0.019\leq\Omega_k\leq 0.0072$~\cite{WMAP7}.

Figures \ref{fig:1dim} and Fig.\ref{fig:fit10} also display clearly
the differences in the results obtained when the full shape
information from the LSS matter power spectrum is included versus when
only the corresponding distance measurement from BAO is accounted
for. We see that, with the expected exception of the neutrino mass (and
correspondingly of $\sigma_8$), both sets of data lead to comparable
precision on the determination of the cosmological
parameters. Concerning the neutrino masses we find that neither of the
two analysis show any evidence for neutrino mass and the best fit
point is obtained for $\sum m_\nu=0$. However the 95 \% upper bound
obtained when using BAO, $\sum m_\nu \leq 1.5$, is tighten by a about a
factor 2, $\sum m_\nu \leq 0.76$, by considering instead the full
LSSPS.
\FIGURE[!h]{
\includegraphics[width=0.9\textwidth]{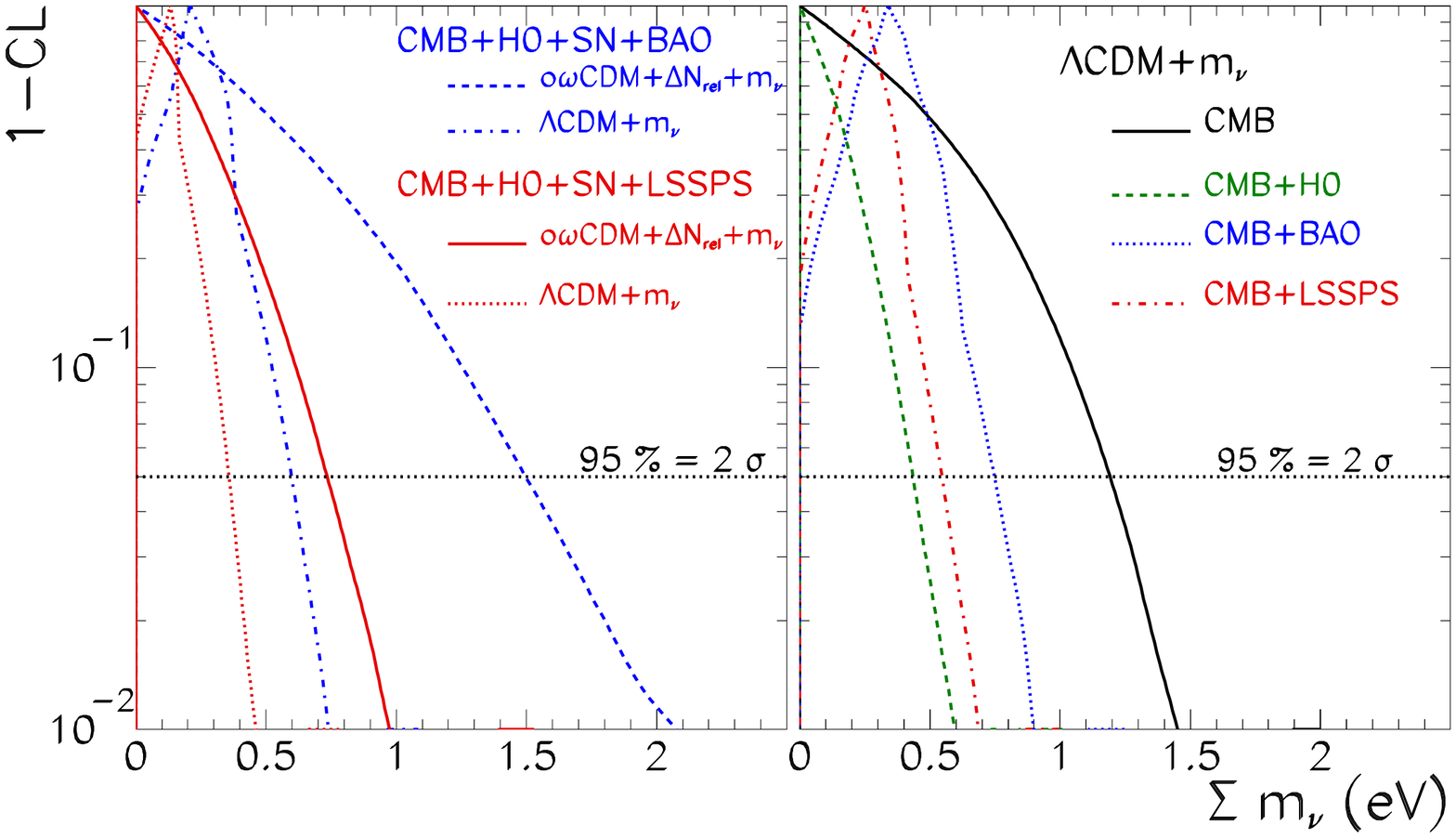}
\caption{\label{fig:mnu} Constraint on $\Sigma m_\nu$ as  a function of the
CL for the different analysis as labeled in the figure.}
}

\TABLE{
\begin{tabular}{|c|c||c|} 
\hline 
Model & Observables & $\Sigma m_\nu$ (eV)   95\% Bound \\ 
\hline
$
o\omega{\rm CDM} 
+\Delta N_{\rm rel}+m_\nu
$ 
& {\footnotesize CMB+HO+SN+BAO} & 
$\leq 1.5$ \\
$
o\omega{\rm CDM} 
+\Delta N_{\rm rel}+m_\nu
$ 
& {\footnotesize CMB+HO+SN+LSSPS} & 
$\leq 0.76$ \\
$\Lambda{\rm CDM} +m_\nu$
& {\footnotesize CMB+H0+SN+BAO} & 
$\leq 0.61$ 
\\
$\Lambda{\rm CDM} +m_\nu$
& {\footnotesize CMB+H0+SN+LSSPS} & 
$\leq 0.36$  \\
$\Lambda{\rm CDM} +m_\nu$
& {\footnotesize CMB (+SN)} & 
$\leq 1.2$ \\
$\Lambda{\rm CDM} +m_\nu$
& {\footnotesize CMB+BAO} & 
$\leq 0.75$  \\
$\Lambda{\rm CDM} +m_\nu$
& {\footnotesize CMB+LSSPS} & 
$\leq 0.55$ \\
$\Lambda{\rm CDM} +m_\nu$
& {\footnotesize CMB+H0} & 
$\leq 0.45$ \\
\hline
\end{tabular}
\caption{95 \% upper bound on the sum of the neutrino masses 
from the different cosmological analysis. 
The analysis
within $\Lambda{\rm CDM} +m_\nu$ models 
including only CMB data or in combination with SN yield the same 95\% bound. } 
\label{tab:sigma}
}

We plot in Fig.\ref{fig:mnu} the bound on  $\sum m_\nu$ for the 
two analysis in the 
$o\omega{\rm CDM}+\Delta N_{\rm rel}+m_\nu$ cosmologies at a given 
CL together with the corresponding results from different analysis 
performed in the framework of $\Lambda{\rm CDM}+m_\nu$ models.
The corresponding 95\% CL bounds are listed in Table.~\ref{tab:sigma}.
We find that for the same combination of observables 
CMB+HO+SN+BAO (CMB+HO+SN+LSSPS) the bound for a  $\Lambda{\rm CDM}+m_\nu$
scenario is  $\sum m_\nu \leq 0.61$  ($\sum m_\nu \leq 0.35$) which is
a factor $\sim 3$ (2) tighter than the corresponding one obtained in 
$o\omega{\rm CDM}+\Delta N_{\rm rel}+m_\nu$ cosmologies.
However, we also find that at lower CL $\Lambda{\rm CDM}+m_\nu$ scenarios
are better fitted with a non vanishing $m_\nu$ when the information 
from CMB (and H0) is combined with the   information from LSS surveys. 
This is, however, at most a 1$\sigma$ effect associated with the 
slight mismatches between the best fit values of the cosmological 
parameters obtained in the analysis of the observables 
in the context of the $\Lambda$CDM. This illustrates the well-known 
fact that overconstrained scenarios are more ``sensitive'' to small
fluctuations in the data, due either to statistics or to an optimistic
estimate of the systematic uncertainties. Consequently, even if 
more conservative, the bounds derived on more general scenarios 
are more robust against these effects.

\section{Combination with Oscillation Data}
\label{sec:mbb}
We present in this section the allowed ranges for  the sum 
of the neutrino masses and the two laboratory probes of the 
absolute scale of neutrino mass: the effective neutrino mass in single 
beta decay $m_{\nu_e}$ and  the effective Majorana neutrino mass 
in neutrinoless  $\beta\beta$ decay $m_{ee}$, obtained 
from the combination of the cosmological analysis discussed above, 
with the information  from the global analysis of solar, atmospheric, 
reactor and accelerator longbaseline (LBL) neutrino experiments 
in terms of flavour  oscillations between the three neutrinos~\cite{ourfit}.

Our starting point is the $\chi^2$ function from the oscillation analysis
\begin{eqnarray}
\chi^2_{\rm O}(\Delta m^2_{21},\Delta m^2_{31},\theta_{12},\theta_{13},
\theta_{23},\delta_{\rm CP})&=& 
\chi^2_{\rm Solar+KamLAND} (\Delta m^2_{21},\theta_{12},\theta_{13})
+\chi^2_{\rm CHOOZ} (\Delta m^2_{31},\theta_{13})  \nonumber\\
&&
+\chi^2_{\rm ATM+LBL} (\Delta m^2_{21},\Delta m^2_{31},\theta_{12},\theta_{13},
\theta_{23},\delta_{\rm CP})   \label{eq:chiosc} \\  
&\Rightarrow&  \chi^2_{\rm O} (m_{\nu_e},m_{ee}, \sum m_{\nu_i})
\end{eqnarray}
where the last step is obtained  after marginalization over 
$\Delta m^2_{31}$ and $\theta_{23}$ and
allowing for variation of the two phases $\eta_1$ and $\eta_2$ within 
their full range.

In Fig.\ref{fig:mbeta} we plot the 95\% allowed regions (for 2 dof) 
in the planes   ($m_{\nu_e}$,$\sum  m_\nu$)  and ($m_{ee}$,$\sum  m_\nu$) 
as obtained from the marginalization of 
$\chi^2_{\rm O} (m_{\nu_e},m_{ee}, \sum m_{\nu_i})$ with respect to the
undisplayed parameter in each plot. In the figure we also show 
superimposed the single parameter 95\%  bounds on 
$\sum m_{\nu_i}$ from the different cosmological analysis described in the
previous section.  The figure illustrates the well-known fact that
currently for either mass ordering the results from neutrino oscillation 
experiments imply a lower bound on $m_{\nu_e}$. On the contrary  $m_{ee}$
is only bounded from below for the case of the normal ordering while 
full cancellation due to the unknown Majorana phases is still allowed
for the inverted ordering. 

\TABLE{
\begin{tabular}{|c|c||c|c|c|} 
\hline
& &  \multicolumn{3}{c|}{Cosmo+Oscillations}\\
 &   &\multicolumn{3}{c|}{95\% Ranges} \\
\hline 
Model & Observables 
&  $m_{\nu_e}$  (eV) & $m_{ee}$  (eV) &  $\Sigma m_\nu$ (eV) \\
\hline
$\begin{array}{c}
o\omega{\rm CDM} \\
+\Delta N_{\rm rel}+m_\nu
\end{array}$ 
& {\footnotesize CMB+HO+SN+BAO} & 
$\begin{array} {l} 
{\rm N} \; [0.0047- 0.51]\\ 
{\rm I}\; [0.047- 0.51] 
\end{array}$ &
$\begin{array} {l} 
{\rm N} \; [0.00- 0.51]\\ 
{\rm I}\; [0.014- 0.51] 
\end{array}$ &
$\begin{array} 
{l} {\rm N} \; [0.056- 1.5]\\ 
{\rm I}\; [0.098- 1.5] 
\end{array}$ \\[+0.5cm]
$\begin{array}{c}
o\omega{\rm CDM} \\
+\Delta N_{\rm rel}+m_\nu
\end{array}$ 
& {\footnotesize CMB+HO+SN+LSSPS} & 
$\begin{array} {l} 
{\rm N} \; [0.0047- 0.27]\\ 
{\rm I}\; [0.047- 0.27] 
\end{array}$ &
$\begin{array} {l} 
{\rm N} \; [0.00- 0.25]\\ 
{\rm I}\; [0.014- 0.25] 
\end{array}$ &
$\begin{array} 
{l} {\rm N} \; [0.056- 0.75]\\ 
{\rm I}\; [0.098- 0.76] 
\end{array}$ \\[+0.5cm]
$\Lambda{\rm CDM} +m_\nu$
& {\footnotesize CMB+H0+SN+BAO} & 
$\begin{array} {l} 
{\rm N} \; [0.0047- 0.20]\\ 
{\rm I}\; [0.048- 0.21] 
\end{array}$ &
$\begin{array} {l} 
{\rm N} \; [0.00- 0.20]\\ 
{\rm I}\; [0.014- 0.21] 
\end{array}$ &
$\begin{array} 
{l} {\rm N} \; [0.056- 0.61]\\ 
{\rm I}\; [0.097- 0.61] 
\end{array}$ \\[+0.5cm]
$\Lambda{\rm CDM} +m_\nu$
& {\footnotesize CMB+H0+SN+LSSSP} & 
$\begin{array} {l}  
{\rm N} \; [0.0047- 0.12]\\ 
{\rm I}\; [0.047- 0.12] 
\end{array}$ &
$\begin{array} {l} 
{\rm N} \; [0.00- 0.12]\\ 
{\rm I}\; [0.014- 0.12] 
\end{array}$ &
$\begin{array} 
{l} {\rm N} \; [0.056- 0.36]\\ 
{\rm I}\; [0.098- 0.36] 
\end{array}$ \\[+0.5cm]
$\Lambda{\rm CDM} +m_\nu$
& {\footnotesize CMB (+SN)} & 
$\begin{array} {l} 
{\rm N} \; [0.0047- 0.40]\\ 
{\rm I}\; [0.047- 0.40] 
\end{array}$ &
$\begin{array} {l} 
{\rm N} \; [0.00- 0.40]\\ 
{\rm I}\; [0.014- 0.41] 
\end{array}$ &
$\begin{array} 
{l} {\rm N} \; [0.056- 1.2]\\ 
{\rm I}\; [0.098- 1.2] 
\end{array}$ \\[+0.5cm]
$\Lambda{\rm CDM} +m_\nu$
& {\footnotesize CMB+BAO} & 
$\begin{array} {l} 
{\rm N} \; [0.0052- 0.25]\\ 
{\rm I}\; [0.047- 0.25] 
\end{array}$ &
$\begin{array} {l} 
{\rm N} \; [0.00- 0.25]\\ 
{\rm I}\; [0.014- 0.25] 
\end{array}$ &
$\begin{array} 
{l} {\rm N} \; [0.056- 0.75]\\ 
{\rm I}\; [0.099- 0.75] 
\end{array}$ \\[+0.5cm]
$\Lambda{\rm CDM} +m_\nu$
& {\footnotesize CMB+LSSPS} & 
$\begin{array} {l} 
{\rm N} \; [0.0047- 0.18]\\ 
{\rm I}\; [0.048- 0.19] 
\end{array}$ &
$\begin{array} {l} 
{\rm N} \; [0.00- 0.18]\\ 
{\rm I}\; [0.014- 0.19] 
\end{array}$ &
$\begin{array} 
{l} {\rm N} \; [0.056- 0.55]\\ 
{\rm I}\; [0.099- 0.55] 
\end{array}$ \\[+0.5cm]
$\Lambda{\rm CDM} +m_\nu$
& {\footnotesize CMB+H0} & 
$\begin{array} {l} 
{\rm N} \; [0.0047- 0.14]\\ 
{\rm I}\; [0.047- 0.16] 
\end{array}$ &
$\begin{array} {l} 
{\rm N} \; [0.00- 0.14]\\ 
{\rm I}\; [0.014- 0.16] 
\end{array}$ &
$\begin{array} 
{l} {\rm N} \; [0.056- 0.44]\\ 
{\rm I}\; [0.097- 0.45] 
\end{array}$ \\[+0.5cm]
\hline
\end{tabular}
\caption{95\% allowed
ranges for the different probes of the absolute neutrino mass
scale  from the global analysis of the cosmological data with 
with the results from oscillation experiments.
The analysis
within $\Lambda{\rm CDM} +m_\nu$ models 
including only CMB data or in combination with SN yield the same 
95\% ranges. } 
\label{tab:mbeta}
}

\FIGURE[!h]{
\includegraphics[width=0.7\textwidth]{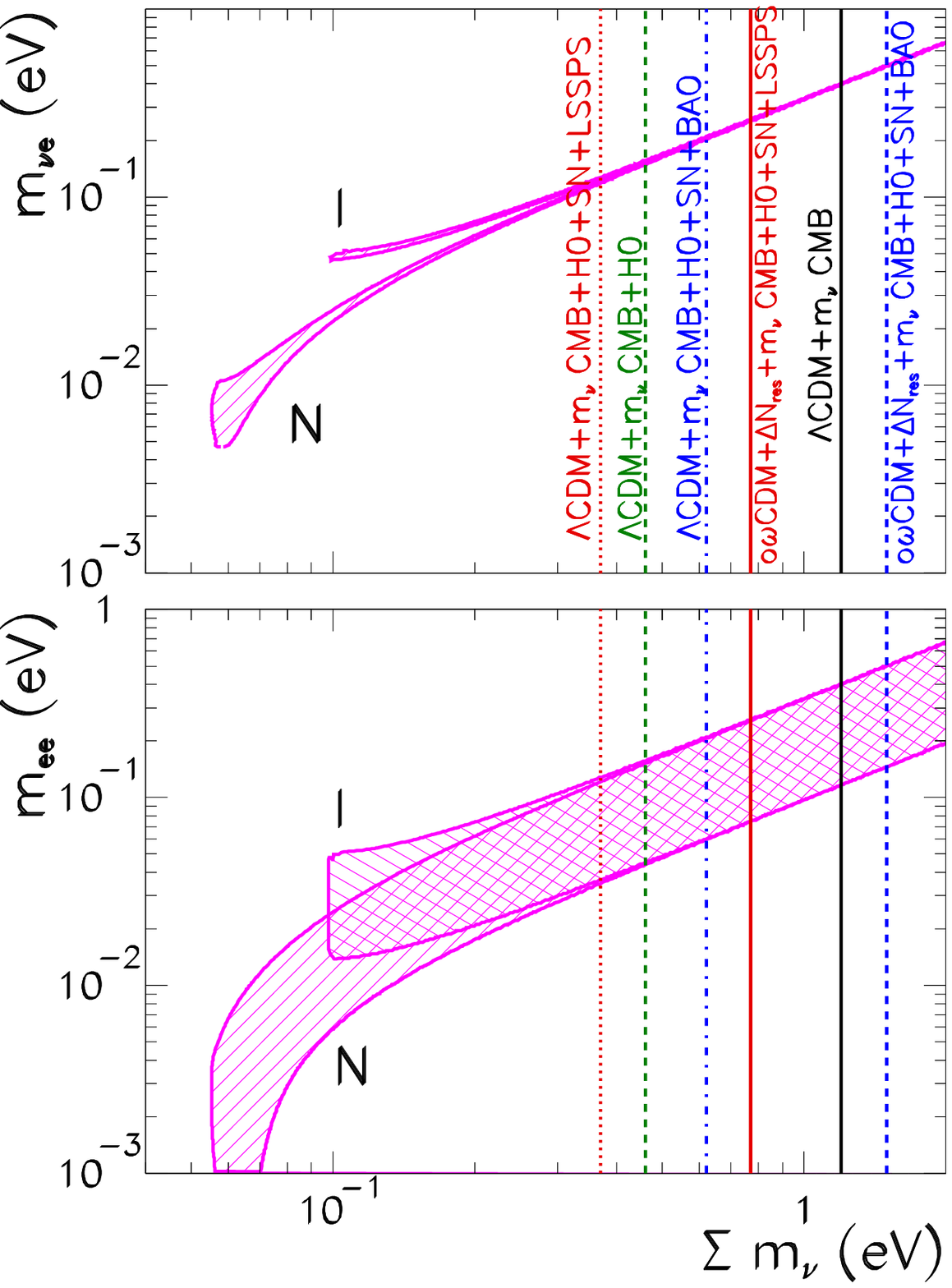}
\caption{
95\% allowed regions (for 2 dof) in the planes  
($m_{\nu_e}$,$\sum  m_\nu$) 
and ($m_{ee}$,$\sum  m_\nu$) from the global analysis of oscillation data
(full regions). We also show 
superimposed the 95\% upper bounds  
on $\sum  m_\nu$ from cosmological constraints
for the different analysis as labeled in the figure.}
\label{fig:mbeta}
}

In order to obtain the global combined ranges we first define a one parameter 
equivalent  $\chi^2_{\rm C} (\sum m_{\nu})$ function~\cite{foglimbeta}
for a given cosmological analysis  
from the condition that it leads to the same CL intervals
than the corresponding 
marginalized one-dimensional probability distribution function: 
\begin{equation}
{\rm CL}
=\frac{1}{2\pi}\int_0^{\chi^2_{\rm C} (\sum m_{\nu})} 
\frac{e^{-x^2/2}}{\sqrt{x}}
\end{equation}
where CL is obtained from Eq.~\eqref{eq:1dimrange} with
$\sum m_{\nu_i}= x^{\rm CL}_{i,\rm max}$ 
and $x^{\rm CL}_{i,\rm min}=0$ when the lower bound for that
CL is $0$  (which imples that the function 
$\chi^2_{\rm C} (\sum m_{\nu})$  
is single valuated).  
When $x^{\rm CL}_{i,\rm min}\neq 0$ the function 
$\chi^2_{\rm C} (\sum m_{\nu})$ takes the same value for 
$\sum m_{\nu_i}= x^{\rm CL}_{i,\rm max}$  and 
$\sum m_{\nu_i}= x^{\rm CL}_{i,\rm min}$. 

Finally we construct 
\begin{equation}
\chi^2_{\rm O+C} (m_{\nu_e},m_{ee}, \sum m_{\nu_i})=
\chi^2_{\rm O} (m_{\nu_e},m_{ee}, \sum m_{\nu_i})+
\chi^2_{\rm C} (\sum m_{\nu}) \;,  
\end{equation}
from which we obtain the the 2$\sigma$ 1-dim allowed ranges
for $m_{\nu_e}$, $m_{ee}$, and $\sum m_{\nu_i}$ given  in 
Table~\ref{tab:mbeta} from the condition 
\begin{equation}
\Delta\chi^2_{\rm O+C}(m_{\nu_e})={\rm Min}_{(m_{ee}, \sum m_{\nu_i})}
\left[ 
\chi^2_{\rm O+C}(m_{\nu_e},m_{ee}, \sum m_{\nu_i})\right]
-\chi^2_{\rm O+C,min}<4 \; , 
\end{equation}
and equivalently for $m_{ee}$ and $\sum m_{\nu_i}$. 

The results show that, even for the most restrictive analysis including  
LSSPS, part of the allowed ranges for $m_{\nu_e}$ in 
the context of the $o\omega{\rm CDM}+\Delta N_{\rm rel}+m_\nu$ 
cosmologies are within the reach of the KATRIN experiment. On the
contrary this is not the case for  $\Lambda{\rm CDM}+m_\nu$ models
unless only the information of CMB and BAO (or SN) is included. 
We also find that near future neutrinoless double beta decay 
can  test some of the allowed ranges in all these scenarios. 
This will be complementary to  the improvement on the 
expected sensitivity from upcoming cosmological probes such as the 
Planck mission \cite{planck}.

\section{Summary}
\label{sec:conclu}
In this work we have studied the information on the absolute value
of the neutrino mass which can be obtained from the analysis of
the cosmological data  in $o\omega{\rm CDM}+\Delta N_{\rm rel}+m_\nu$ 
cosmologies in which 
besides neutrino masses, one allows for non-vanishing curvature, 
dark energy  with equation of state with $\omega\neq -1$ together with
the presence of new particle physics whose effect on the present 
cosmological observations can be  parametrized in terms of additional 
relativistic degrees of freedom. To break the degeneracies in these
models, at least the information from four different cosmological 
probes must be combined.  Thus we have performed analysis 
including the data from  CMB  experiments, the present day Hubble constant H0, 
measurement,  the high-redshift Type-I SN results and the 
information from large scale LSS surveys. We have compared 
the results from the analysis when the full shape information from the LSS
matter power spectrum is included versus when only the corresponding
distance measurement from the baryon acoustic oscillations is consider.

Our results are summarize in Table \ref{tab:1dim}.
Because of the degeneracies present in these cosmologies one finds 
a degradation in the constraints of the {\sl standard}
parameters (ie those of the $\Lambda$CDM model) when compared with the
analysis performed within the $\Lambda$CDM priors for the same set of
observables. Concerning the neutrino masses we find that neither of the
two analysis show any evidence for neutrino mass and the best fit 
is obtained for $\sum m_\nu=0$. However the 95 \% upper bound
obtained when using BAO, $\sum m_\nu \leq 1.5$, is tighten by a about a
factor 2, $\sum m_\nu \leq 0.76$, by considering instead the full
LSSPS. We have compared these results with those obtained 
from different analysis  performed in the framework of 
$\Lambda{\rm CDM}+m_\nu$ models.
The corresponding 95\% CL bounds are listed in Table.~\ref{tab:sigma}.
We find that for the same combination of observables 
CMB+HO+SN+BAO (CMB+HO+SN+LSSPS) the bound for a  $\Lambda{\rm CDM}+m_\nu$
scenario is  $\sum m_\nu \leq 0.61$  ($\sum m_\nu \leq 0.35$) which is
a factor $\sim 3$ (2) tighter than the corresponding one obtained in 
$o\omega{\rm CDM}+\Delta N_{\rm rel}+m_\nu$ cosmologies.

Finally we have statistically combined these results with the 
information on neutrino mass differences and mixing from the global
analysis of neutrino oscillation experiments and we have derived the
presently allowed ranges for the two laboratory probes of the absolute
scale of neutrino mass: the effective neutrino mass in single beta
decay $m_{\nu_e}$ and the effective Majorana neutrino mass in
neutrinoless $\beta\beta$ decay $m_{ee}$. These results can be used to 
directly address the capabilities of the future $\beta$ and 
neutrinoless-$\beta\beta$ decay experiments to probe the allowed 
parameter space.

\section*{Acknowledgments}

We are specially indebted to  J. Taron for his collaboration in the
early stages of this project and comments on the final version. 
We thank R. Jimenez, E. Nardi and L. Verde for comments. 
This work is supported by Spanish MICINN grants 2007-66665-C02-01,
FPA-2009-08958 and FPA-2009-09017 and consolider-ingenio 2010 grant
CSD-2008-0037, by CSIC grant 200950I111,
by CUR Generalitat de Catalunya grant 2009SGR502, by
Comunidad Autonoma de Madrid through the HEPHACOS project P-ESP-00346,
by USA-NSF grant PHY-0653342 and by EU grant EURONU.

\bibliographystyle{JHEP}
\providecommand{\href}[2]{#2}\begingroup\raggedright
\endgroup
\end{document}